\documentclass{jfm}

\usepackage{graphicx}
\usepackage{natbib}
\usepackage{color}

\ifCUPmtlplainloaded \else
  \checkfont{eurm10}
  \iffontfound
    \IfFileExists{upmath.sty}
      {\typeout{^^JFound AMS Euler Roman fonts on the system,
                   using the 'upmath' package.^^J}%
       \usepackage{upmath}}
      {\typeout{^^JFound AMS Euler Roman fonts on the system, but you
                   dont seem to have the}%
       \typeout{'upmath' package installed. JFM.cls can take advantage
                 of these fonts,^^Jif you use 'upmath' package.^^J}%
      }
  \else
  \fi
\fi

\ifCUPmtlplainloaded \else
  \checkfont{msam10}
  \iffontfound
    \IfFileExists{amssymb.sty}
      {\typeout{^^JFound AMS Symbol fonts on the system, using the
                'amssymb' package.^^J}%
       \usepackage{amssymb}%

      }{}
  \fi
\fi

\ifCUPmtlplainloaded \else
  \IfFileExists{amsbsy.sty}
    {\typeout{^^JFound the 'amsbsy' package on the system, using it.^^J}%
     \usepackage{amsbsy}}
    {\providecommand\boldsymbol[1]{\mbox{\boldmath $##1$}}}
\fi

\newsavebox{\astrutbox}
\sbox{\astrutbox}{\rule[-5pt]{0pt}{20pt}}

\title[Reflexions on dissipation]{Reflections on dissipation associated with thermal convection}

\author[T. Alboussi\`ere and Y. Ricard]%
{Thierry Alboussi\`ere
  \thanks{Email address for correspondence: thierry.alboussiere@ens-lyon.fr}
and Yanick Ricard}

\affiliation{Laboratoire de G\'eologie de Lyon, CNRS, ENS-Lyon, Universit\'e Lyon 1, France}

\pubyear{2013}
\volume{725}

\date{10 December 2012; revised 11 March 2013; accepted 8 May 2013}

\begin{document}

\centerline{\bf \LARGE Reflections on dissipation associated with }
\centerline{\bf \LARGE thermal convection}

\vspace*{5 mm}

\centerline{Thierry Alboussi\`ere and Yanick Ricard}

\vspace*{5 mm}

\centerline{\small \rm Laboratoire de G\'eologie de Lyon, UMR 5276, CNRS, ENS-Lyon}
\centerline{\small \rm Universit\'e Lyon 1, 69622 Villeurbanne, France
}

\vspace*{5 mm}

\centerline{3 November 2019}

\vspace*{15 mm}


\begin{minipage}{13 cm}{
		This paper was originally published in June 2013 in JFM (vol. 725). There was one result concerning an expression of viscous dissipation in convective flows and a second result on the condition of applicability of the anelastic liquid equation ($\gamma -1 \ll 1$). We then found that this second result was wrong and the explanation why this is so was published as a corrigendum in July 2014 in JFM (vol. 751). In a few words, the origin of the error is that we have derived an expression for dissipation in the framework of the exact equations and subsequently used it in the framework of the anelastic models. Here we have included the original article, followed by its corrigendum.}
\end{minipage}

\thispagestyle{empty}

\newpage

\setcounter{page}{1}

\maketitle

\begin{abstract}
Buoyancy-driven convection is modelled using the Navier-Stokes and entropy equations. It is first shown that the coefficient of heat capacity at constant pressure, $c_p$, must in general depend explicitly on pressure ({\it i.e.} is not a function of temperature alone) in order to resolve a dissipation inconsistency. It is shown that energy dissipation in a statistically steady state is the time-averaged volume integral of $ - \frac{\mathrm{D} P}{\mathrm{D} t} $ and not that of $ - \alpha T \frac{\mathrm{D} P}{\mathrm{D} t} $. Secondly, in the framework of the anelastic equations derived with respect to the adiabatic reference state, we obtain a condition when the anelastic liquid approximation can be made, $ \gamma -1  \ll 1 $, independent of the dissipation number. 
\end{abstract}

\noindent
	{\sl keywords: compressible convection, thermal convection, dissipation, anelastic approximation, anelastic liquid approximation}

\section{The general expression of dissipation}
Convection in the geophysical context is often associated with compressible effects (as early as in the work of \cite{carnot}) and simultaneously with a significant energy dissipation \citep[see][]{backus}, of the same order of magnitude as the heat flux. In the case of convection in the Earth's mantle, it is important to determine the distribution of dissipation as it is strongly coupled with effective viscosity and plays a role in the structure of convection \citep[see][]{berco96,tackley96}, while the amount of dissipation in the Earth's core is a direct measure of the energy available to the geodynamo \citep[see][]{Christensen_Tilgner04,buffett02}. In this paper we shall be interested specifically in dissipation in 'liquids', defined as those fluids with a small product $\alpha T$ compared to unity, where $\alpha$ is the coefficient of thermal expansion at constant pressure and $T$ is the thermodynamic temperature. This product is unity for perfect and semi-perfect gases, but is usually very small for liquids, of order $0.05$ both in the outer core of the Earth, and in the mantle which are hence both usually considered as 'liquids' from the perspective of convection.  

Let us consider a simple configuration of a region of Newtonian fluid with no internal heat production, bounded with solid walls or stress-free boundaries  and prescribed temperatures or heat fluxes within a uniform, constant gravity field ${\bf g}$. It might be a simple Rayleigh-B\'enard configuration with an imposed temperature difference. It might also be more complex, we simply require that the boundary exerts no work on the fluid region and that a statistically steady state can be reached so that the net heat input is zero. Density $\rho$, velocity ${\bf u}$, pressure $P$, temperature $T$ and entropy $s$ satisfy the continuity, Navier-Stokes and entropy (or heat transfer) equations  
\begin{eqnarray}
\frac{\partial \rho }{\partial t } + {\bf \nabla} \cdot \left( \rho {\bf u} \right) &=& 0, \label{continuity} \\
\rho \frac{\mathrm{D} {\bf u} }{\mathrm{D} t } &=& - {\bf \nabla } P + \rho {\bf g} + {\bf \nabla } \cdot {\bf \tau }, \label{NS} \\
\rho T  \frac{\mathrm{D} s }{\mathrm{D} t } &=&   {\bf \dot \epsilon } \colon {\bf \tau }  - {\bf \nabla } \cdot {\bf \phi }  , \label{entropy} 
\end{eqnarray}
where ${\bf \phi } = - k {\bf \nabla } T $ is the conduction heat flux (with thermal conductivity $k$), ${\bf \tau} $ is the stress tensor and ${\bf \dot \epsilon}$ is the tensor of the rate of deformation.
Rewriting (\ref{entropy}) using the thermodynamic relation $T \mathrm{d} s = c_p \mathrm{d} T - \frac{\alpha T}{\rho} \mathrm{d} P$ (the well-known equation (\ref{dS}) re-derived in the appendix), with $c_p$ the specific heat capacity at constant pressure, leads to  
\begin{equation}
\rho c_p \frac{\mathrm{D} T }{\mathrm{D} t } - \alpha T \frac{\mathrm{D} P }{\mathrm{D} t }  =  {\bf \dot \epsilon } \colon {\bf \tau }  - {\bf \nabla } \cdot {\bf \phi }   . \label{heat} 
\end{equation}
A classical expression for the energy dissipation is obtained by integration of equation (\ref{heat}) over the whole fluid domain, under the tentative assumption that $c_p$ can be taken as uniform, and by taking its time-average \citep[see][]{mckenzie_jarvis,hewitt_etc}:   
\begin{equation}
\left< {\bf \dot \epsilon } \colon {\bf \tau } \right> = - \left< \alpha T \frac{\mathrm{D} P }{\mathrm{D} t } \right>, \label{Hewitt85}
\end{equation}
where $\left< \cdot \right>$ denotes time-averaged volume integral. However, the integration of the dot product of the Navier-Stokes equation (\ref{NS}) with velocity ${\bf u}$, assuming no work is done through the boundary, leads to
\begin{equation}
\left< {\bf \dot \epsilon } \colon {\bf \tau } \right> = - \left<  {\bf u} \cdot {\bf \nabla } P \right> = - \left< \frac{\mathrm{D} P }{\mathrm{D} t } \right>, \label{correct_dissip}
\end{equation}
which is very different from equation (\ref{Hewitt85}) for liquids as $\alpha T \ll 1$. The inconsistency is resolved when it is recognized that $c_p$ is generally not uniform nor a function of temperature only. As $c_p = \left. \frac{\partial H}{\partial T} \right| _P  $ by definition, with $H$ the enthalpy, the term $\rho c_p \frac{\mathrm{D} T }{\mathrm{D} t }$ in equation (\ref{heat}) can be integrated as follows
\begin{equation}
\left< \rho c_p \frac{\mathrm{D} T }{\mathrm{D} t } \right> = \left< \rho \frac{\mathrm{D} H }{\mathrm{D} t } \right> - \left< \rho \left. \frac{\partial H}{\partial P} \right| _T  \frac{\mathrm{D} P }{\mathrm{D} t } \right> . \label{integ}
\end{equation}
The first term of the right-hand side is always zero (this would be true for any function of state, not just for $H$) in the statistically steady case. The second term is expressed using equation (\ref{dHdP}) of the appendix, so that 
\begin{equation}
\left< \rho c_p \frac{\mathrm{D} T }{\mathrm{D} t } \right> = - \left< \left( 1 - \alpha T   \right) \frac{\mathrm{D} P }{\mathrm{D} t } \right> . \label{interm}
\end{equation}
By making the hypothesis of a uniform $c_p$, an important term has been forgotten, thus invalidating expression (\ref{Hewitt85}).
The correct averaging of the heat equation, just like the balance of mechanical energy, leads to (\ref{correct_dissip}).
What we have done is to show that it does not make sense to assume $c_p$ constant when $\alpha T \ll 1$. Equation (\ref{dHdP}) imposes that $H$ depends on pressure, that dependence cannot be made as weak as one wishes and has consequences on the term of the dissipation budget involving $c_p$ treated in equation (\ref{integ}). 

Incidently, we may just write two other exact expressions for dissipation, at least as fundamental as (\ref{correct_dissip}), obtained from the integration of (\ref{heat}) where entropy variations are not expanded in terms of pressure and temperature variations (see \cite{verhoogen80}, page 82):
\begin{equation}
\left< \tau \colon \dot \epsilon  \right> = \left< \rho \, T \, \frac{\mathrm{D} s }{\mathrm{D} t } \right> = - \left< \rho \, s \, \frac{\mathrm{D} T }{\mathrm{D} t } \right> . \label{two_more}
\end{equation}

\section{Dissipation in the anelastic approximation (AA)}
Equation (\ref{correct_dissip}) will now be used, within the anelastic approximation (AA) \citep[see][]{ogura_phillips}, to determine a condition when the anelastic liquid approximation (ALA) \citep[see][]{braginsky_roberts95,schubert_etc} can be made. We assume that convection is sufficiently vigorous to drive the fluid close to an iso-entropy state, commonly called 'adiabatic'. All quantities will be expanded around their depth-dependent adiabatic reference state denoted with the subscript $0$ while fluctuations are indicated by primes. The reference state is dependent on height only and is defined such that the entropy $s_0$ is uniform and that the hydrostatic equation is satisfied $\partial P_0 / \partial z = - \rho _0 g$, where $z$ is the vertical upward coordinate. Equation (\ref{dS}) shows that, when entropy is vigorously mixed, a relation develops between pressure and temperature variations. In addition, when hydrostatics is considered, this leads to the so-called 'adiabatic gradient'
\begin{equation}
\frac{d T_0}{d z} = - \frac{\alpha _0 g T_0}{c_{p0}} . \label{grad_adiab}
\end{equation} 
The main purpose of the anelastic approximation is to eliminate sound waves by replacing the general continuity equation (\ref{continuity}), by its zeroth-order expansion
\begin{equation}
{\bf \nabla } \cdot \left( \rho _0 {\bf u} \right) = 0. \label{continuity0}
\end{equation}
The first order expansion of the entropy equation (\ref{entropy}) is
\begin{equation}
\rho _0 T_0 \frac{\mathrm{D} s' }{\mathrm{D} t } =  {\bf \dot \epsilon } \colon {\bf \tau }  - {\bf \nabla } \cdot \left( \phi _0 + \phi ' \right)  . \label{entropy_lin_0} 
\end{equation}
Writing the first term in a conservative form introduces an advection term for the adiabatic temperature profile which is expressed using the adiabatic gradient (first term on the right-hand side of the following equation) 
\begin{equation}
\rho _0 \frac{\mathrm{D} \left( T_0 s' \right) }{\mathrm{D} t } = - \frac{\alpha _0 \rho _0 T_0 g}{c_{p0}} u_z s'  +  {\bf \dot \epsilon } \colon {\bf \tau }  - {\bf \nabla } \cdot \left( \phi _0 + \phi ' \right)  . \label{entropy_lin} 
\end{equation}
The derivation of the anelastic momentum equation is then based on the expansion of density fluctuations in terms of entropy and pressure fluctuations
\begin{equation}
\rho ' = \left. \frac{\partial \rho _0}{\partial s} \right| _P s' + \left. \frac{\partial \rho _0}{\partial P} \right| _s P'. \label{expan_rho}
\end{equation}
Using the Maxwell relation associated with (\ref{GibbsH}) and the expression for the adiabatic temperature gradient obtained from (\ref{dS}) for the first term, and hydrostatics for the second, one obtains
\begin{equation}
\rho ' =  - \frac{\alpha _0 \rho _0 T_0}{c_{p0}} s' - \frac{1}{ \rho _0 g } \frac{\partial \rho _0}{\partial z} P'. \label{expan_rho_bis}
\end{equation}
Substituting in the Navier-Stokes equation (\ref{NS}) and selecting the lowest order terms leads to
\begin{equation}
\rho _0 \frac{\mathrm{D} {\bf u} }{\mathrm{D} t } = - {\bf \nabla } P' + \frac{\alpha _0 \rho _0 T_0 g}{c_{p0}} s' {\hat {\bf e}_z} + \frac{1}{\rho _0} \frac{\partial \rho _0}{\partial z} P' {\hat {\bf e}_z} + {\bf \nabla } \cdot {\bf \tau }, \label{NS_lin} 
\end{equation}
where ${\hat {\bf e}_z}$ is the vertical unit vector. The first and third terms on the right-hand side can be put together in a single conservative term
\begin{equation}
\rho _0 \frac{\mathrm{D} {\bf u} }{\mathrm{D} t } = - \rho _0 {\bf \nabla } \left( \frac{P'}{\rho _0 }  \right) + \frac{\alpha _0 \rho _0 T_0 g}{c_{p0}} s' {\hat {\bf e}_z} + {\bf \nabla } \cdot {\bf \tau }. \label{NS_lin_bis} 
\end{equation}
This form of the dynamic equation was derived by \cite{braginsky_roberts95}. 
From this point, we can derive important consequences in terms of energy dissipation. 
In (\ref{NS_lin_bis}) the work of $-\rho _0 {\bf \nabla } \left( {P'}/{\rho _0 }  \right)$ integrated over the volume is zero. Therefore the mechanical works of the first ($-{\bf \nabla }P'$) and third ($P'{\bf \nabla }\rho_0/\rho_0$) terms of (\ref{NS_lin}) are opposite to each other. As the
expression of the dissipation $- \left< \mathrm{D} P / \mathrm{D} t \right>$ becomes $ - \left< \mathrm{D} P' / \mathrm{D} t \right> = - \left< {\bf u} \cdot {\bf \nabla} P' \right>  $, we have
\begin{equation}
\left< {\bf \dot \epsilon } \colon {\bf \tau } \right> =  - \left<   \frac{1}{\rho _0} \frac{\partial \rho _0}{\partial z}  u_z P' \right> = \left<  \frac{ \rho _0 g }{K_{s0}} u_z P' \right> , \label{otherDiss}
\end{equation}
where $K_s$ is the incompressibility 
 at constant entropy.
 Introducing  $c =\sqrt{K_s/\rho}$, the celerity of sound waves, one can finally express the energy dissipation in terms of the correlation of velocity and pressure fluctuations
\begin{equation}
\left< {\bf \dot \epsilon } \colon {\bf \tau } \right> = \left<  \frac{g}{c_0^2} u_z P' \right>. \label{otherDissBis}
\end{equation}
However, energy dissipation must also be equal to the work done by the second term on the right-hand side of (\ref{NS_lin}), in agreement with the expression of dissipation associated with (\ref{entropy_lin}) and the general expression (\ref{two_more}), 
\begin{equation}
\left< {\bf \dot \epsilon } \colon {\bf \tau } \right> = \left< \frac{\alpha _0 \rho _0 T_0 g}{c_{p0}}  u_z s' \right>, \label{otherDissTer}
\end{equation}
now in terms of the correlation between velocity and entropy fluctuations. Expanding entropy fluctuations in terms of temperature and pressure fluctuations, using (\ref{dS}), leads to
\begin{equation}
\left< {\bf \dot \epsilon } \colon {\bf \tau } \right> = \left< \alpha _0 \rho _0 g  u_z T' \right> - \left< \frac{\alpha _0^2 T_0 g}{c_{p0}}  u_z P' \right>.  \label{otherDissTP}
\end{equation}

\section{Energetic validation of the anelastic liquid approximation (ALA)}
Comparing equations (\ref{otherDissBis}) and (\ref{otherDissTP}) shows the relative importance of the pressure and the temperature terms in (\ref{otherDissTP}). The coefficient within the pressure/velocity correlation term can be written  
\begin{equation}
\frac{\alpha _0^2 T_0 g}{c_{p0}}  = \frac{\alpha _0^2 c_0^2 T_0 }{c_{p0}} \frac{g}{c_0^2}  =  (\gamma _0 -1 ) \frac{g}{c_0^2}   , \label{boundP}
\end{equation}
where $\gamma = c_p / c_v $ is the ratio of specific heat capacities at constant pressure and at constant volume. The second equality in (\ref{boundP}) can be retrieved from the general Mayer relation $ c_p - c_v = - T/ \rho ^2 (\partial P / \partial T )_{\left| \rho \right.} (\partial \rho / \partial T )_{\left| P \right.} $.  
In the case of a constant value of $\gamma$, independent of temperature and pressure (one must here verify that there exist equations of state with constant ratio of specific heat coefficients: this is indeed the case for the class of ideal gases, with arbitrary polytropic index), 
equations (\ref{boundP}) and (\ref{otherDissBis}) lead immediately to the fraction of pressure/velocity correlations to dissipation in (\ref{otherDissTP})
\begin{equation}
\left< \frac{\alpha _0^2 T_0 g}{c_{p0}}  u_z P' \right> = ( \gamma _0 -1 )  \left< {\bf \dot \epsilon } \colon {\bf \tau } \right>, \label{boundP2}
\end{equation}
It is hence safe from the point of view of energy dissipation to express entropy fluctuations $s'$ (see \ref{otherDissTer}) in terms of temperature fluctuations only $T'$ (see \ref{boundP2}) in equation (\ref{NS_lin_bis}) under the condition 
\begin{equation}
 \gamma - 1  = \frac{c^2 \alpha ^2 T }{c_{p} } \ll 1. \label{conditionliquid}
\end{equation}
While that condition was obtained rigorously only in the case of constant $\gamma$, we expect it to apply in the general case. 
Furthermore, provided there are reasonable correlations between $u_z$ and $P'$ as well as between  $u_z$ and $T'$, one can expect the condition (\ref{conditionliquid}) to be the correct condition of application of the anelastic liquid approximation. It then follows that, 
  while entropy fluctuations are replaced by temperature fluctuations in (\ref{NS_lin_bis}),  $c_{p0} T'$ must be substituted to $T_0 s'$ in (\ref{entropy_lin}) to ensure a consistent energy balance. 
This leads to
 the following anelastic liquid approximation equations, derived from the anelastic approximation (\ref{continuity0}), (\ref{entropy_lin}) and (\ref{NS_lin_bis}): 
\begin{eqnarray}
{\bf \nabla } \cdot \left( \rho _0 {\bf u} \right) \!\!  &=& \!\!  0, \label{continuity_ala} \\
\rho _0 \frac{D \left( c_{p0} T' \right) }{D t } \!\! &=& \!\! - \alpha _0 \rho _0 g u_z T'  +  {\bf \dot \epsilon } \colon {\bf \tau }  - {\bf \nabla } \cdot \left( \phi _0 + \phi ' \right)  ,  \ \ \ \ \ \label{entropy_ala} \\
\rho _0 \frac{D {\bf u} }{D t } \!\! &=& \!\! - \rho _0 {\bf \nabla } \left( \frac{P'}{\rho _0 }  \right) + \alpha _0 \rho _0 g T' {\hat {\bf e}_z} + {\bf \nabla } \cdot {\bf \tau }. \label{NS_ala} 
\end{eqnarray}
Let us mention here, for completeness, that a so-called truncated anelastic liquid approximation (TALA) has been used (for instance in \cite{tan_gurnis} and \cite{jarvis_mckenzie}) in which the contribution of pressure fluctuations to density fluctuations are also neglected in equation (\ref{expan_rho_bis}). As a consequence, the term $ - \rho _0 {\bf \nabla } \left( {P'} / {\rho _0 }  \right) $ in equation (\ref{NS_ala}) above is changed for $-  {\bf \nabla } P' $. However, as pointed out in \cite{leng_zhong}, this change introduces an imbalance between energy dissipation calculated from the dynamical equation and heat dissipation in the thermal equation. Such a TALA formulation should be avoided when the calculation of energy dissipation is an issue. 

Our condition (\ref{conditionliquid}) is in contrast to a scaling law usually derived \citep[see][]{braginsky_roberts95,anufriev_etc,SpiegelVeronis} from the anelastic liquid approximation momentum equation (\ref{NS_ala}) expressing the typical magnitude of pressure fluctuations as a function of temperature fluctuations, $P' \sim \alpha _0 \rho _0 g L  T'$, 
where $L$ is the typical length-scale of the fluid domain. This leads to the following relation
\begin{equation}
\frac{\alpha _0^2 T_0 g}{c_{p0}}  u_z P' \sim \left( \alpha _0 T_0 \right) \frac{\alpha _0 g L}{c_{p0}} \left[\alpha_0 \rho _0 g u_z T' \right], \label{estimP2}
\end{equation}
which would imply that the anelastic liquid approximation is valid when $(\alpha _0 T_0 ) D \ll 1$, where the dissipation number is $D = \alpha_0 g L / c_{p0}$. This heuristic scaling law (\ref{estimP2}) involving $D$, hence $L$ and $g$, is not compatible with the exact relation (\ref{boundP2}) from which the condition (\ref{conditionliquid}) was obtained solely in terms of fluid properties. Actually, from equations (\ref{otherDissBis}) and (\ref{otherDissTP}), one can infer a more adequate scaling of the pressure fluctuations than $P' \sim \alpha _0 \rho _0 g L  T'$, which is valid in the general anelastic approximation (liquid or not) 
\begin{equation}
P' \sim K_{t0} \alpha _0 T', \label{estimPnew} 
\end{equation}
where $K_t=K_s / \gamma$ is the incompressibility at constant temperature. 

\section{Discussion}
Let us now discuss the implications of the results derived in this paper. 
Concerning the 'exact' model (\ref{continuity}), (\ref{NS}) and (\ref{entropy}), we have shown that it is crucial to use thermodynamically consistent thermo-physical coefficients. Specifically, considering a fluid with constant heat capacity $c_p$ and small product $\alpha T$ of temperature by thermal expansion coefficient leads to a severe inconsistency when determining the energetic dissipation. Whatever $\alpha T$, dissipation is always equal to $\left< - D P / D t \right> $. Once compressibility effects have been taken into account in an anelastic model, no energetic inconsistency can arise even in the case when constant uniform values for $c_{p0}$ and other coefficients are used in equations (\ref{entropy_lin}) and (\ref{NS_lin_bis}). Using the correct expression for dissipation, within the model of the anelastic approximation, we have then been able to obtain the condition $\gamma -1 \ll 1 $ for the anelastic liquid approximation to be valid. 
 That condition is only dependent on materials properties, not on gravity or length-scale. Note that the condition (\ref{conditionliquid}) can be expressed with the Gr\"uneisen parameter, $\Gamma = \alpha K_s / ( \rho c_p )$ used in Solid state Physics, as $c^2 \alpha ^2 T / c_p = \gamma -1 = \Gamma \alpha T$. The Gr\"uneisen parameter is close to unity in the mantle and in the Earth's core and so is the ratio of heat capacities. This is the reason why the condition (\ref{conditionliquid}) is close to $\alpha T \ll 1$ in practice concerning the dynamics of the deep Earth. 
For the Earth's liquid outer core and for the mantle, the dimensionless number in (\ref{conditionliquid}) is of order 0.05 justifying the anelastic liquid approximation. 

From the correct expression for dissipation (\ref{correct_dissip}), we have obtained equations (\ref{entropy_ala}) and (\ref{NS_ala}), in the anelastic liquid approximation, which both lead to the following expression for dissipation
\begin{equation}
\left< {\bf \dot \epsilon } \colon {\bf \tau } \right> = \left< \alpha _0 \rho _0 g T' u_z \right> . \label{work_buoy}
\end{equation} 
Surprisingly, \cite{hewitt_etc} reach the same conclusion when starting from their expression (\ref{Hewitt85}):
\begin{eqnarray}
\left< {\bf \dot \epsilon } \colon {\bf \tau } \right> &=& - \left< \alpha T {\bf u} \cdot {\bf \nabla } P \right> \simeq  - \left< \alpha T \rho _0 g u_z \right> \nonumber \\
  & \simeq & \left< \alpha _0 T' \rho _0 g u_z \right> . \label{Hewitt2}
\end{eqnarray}
This expression is indeed approximately correct for liquids and the apparently very different expressions (\ref{correct_dissip}) and (\ref{Hewitt85}) are nearly equal in the anelastic liquid approximation  $\left<  D P  / D t \right>\approx\left< \alpha T D P / D t  \right>$.  
 This is due to the large cancellations in (\ref{correct_dissip}) which allow the same result to be reached when multiplied by a small term $\alpha T$, with appropriate fluctuations (here $\alpha _0 T'$ to the leading order). We only expect significantly different values from expressions (\ref{correct_dissip}) and (\ref{Hewitt85}) when $\alpha T$ is 
 neither too small nor too close to unity, where (\ref{correct_dissip}) and (\ref{Hewitt85}) become similar. 

To conclude, we hope that we have contributed to a clarification of the expression of dissipation in a convective system and of the anelastic liquid approximation. We have not systematically investigated the consequences of these findings in the different possible fields of application: dynamics of the Earth's core and mantle, dynamics of giant planets, ice planets, super-Earth exoplanets. Regarding the application of the anelastic liquid approximation for the terrestrial planets obeying a generic Murnaghan equation of state \citep[see][]{murnaghan},  
 we have no indication that the new criterion (\ref{conditionliquid}), $\gamma -1 \ll 1 $, differs significantly from the classical criterion $\alpha T D \ll 1$. According to Murnaghan's equation of state, the coefficient of expansion $\alpha$ decreases strongly when density $\rho$ is increased, $\alpha \sim \rho ^{-n} $ while the coefficient of incompressiblity increases with $\rho $, $K_s \sim \rho ^n$, with a value of $n$ around 3. We have considered the case of silicate planets from one up to possibly ten Earth masses. The parameter $\gamma -1$ is always less than 0.08, and that value decreases very quickly away from the surface to much lower values in the bulk of the mantle (decreasing also when planetary mass increases). Provided the general anelastic approximation can be applied to mantle dynamics, the anelastic liquid approximation is indeed well justified for the mantle dynamics of the Earth and even better justified for super-Earths.

\begin{appendix}

\section{Useful thermodynamic relations}

From the Gibbs equation expressed in terms of the Gibbs free energy $G$
\begin{equation}
\mathrm{d}G = - S \mathrm{d} T + \frac{1}{\rho } \mathrm{d}P, \label{Gibbs}
\end{equation}
the following Maxwell relation is derived, expressing the partial derivative of entropy with respect to pressure at constant temperature
\begin{equation}
\left.  \frac{\partial S}{\partial P } \right| _T   = - \left.  \frac{\partial V}{\partial T } \right| _P = - \frac{\alpha }{\rho } .  \label{dSdP}
\end{equation}
The partial derivative of entropy with respect to temperature at constant pressure is obtained as follows. Two expression for $dH$ are first written, one from Gibbs relation, the other from the definition of $c_p$
\begin{eqnarray}
\mathrm{d} H & = & T \mathrm{d}S + \frac{1}{\rho } \mathrm{d}P, \label{GibbsH} \\
\mathrm{d} H & = & c_p \mathrm{d}T + \left.  \frac{\partial H}{\partial P } \right| _T \mathrm{d}P. \label{defcp} 
\end{eqnarray}
The difference of these expressions above leads to
\begin{equation}
T \mathrm{d}S = c_p \mathrm{d}T + \left( \left.  \frac{\partial H}{\partial P } \right| _T - \frac{1}{\rho }   \right) \mathrm{d}P . \label{TdS} \\
\end{equation}
This shows that the partial derivative of entropy with respect to temperature at constant pressure is $c_p / T$. Along with equation (\ref{dSdP}), this leads to
\begin{equation}
\mathrm{d}S = \frac{c_p }{T} \mathrm{d}T - \frac{\alpha }{\rho } \mathrm{d}P . \label{dS} \\
\end{equation}
Comparing equations (\ref{TdS}) and (\ref{dS}) leads to an important expression used in this paper
\begin{equation}
\left.  \frac{\partial H}{\partial P } \right| _T = \frac{1 - \alpha T}{\rho }. \label{dHdP}
\end{equation}
The partial derivative of this last equation with respect to temperature at constant pressure leads to expressions of the derivative of $c_p$ with respect to pressure at constant temperature
\begin{eqnarray}
\left.  \frac{\partial  c_p }{\partial P } \right| _T &=& \left. \frac{\partial }{\partial T} \left( \frac{1 - \alpha T}{\rho } \right) \right| _P , \label{dcpdP} \\ 
&=& - \frac{T}{\rho } \left[ \alpha ^2 + \left. \frac{\partial \alpha }{\partial T}  \right| _P  \right] . \label{dcpdPbis}
\end{eqnarray}

\end{appendix}


\newpage

\setcounter{page}{1}
\setcounter{equation}{0}
\setcounter{section}{0}
 
\centerline{\bf \LARGE Reflections on dissipation associated with }
\centerline{\bf \LARGE thermal convection -- CORRIGENDUM}

\vspace*{5 mm}

\centerline{Thierry Alboussi\`ere and Yanick Ricard}

\vspace*{5 mm}

\centerline{\small \rm Laboratoire de G\'eologie de Lyon, UMR 5276, CNRS, ENS-Lyon, Universit\'e Lyon 1,France}

\vspace*{5 mm}

\centerline{July 2014}

\vspace*{5 mm}

\noindent
	{\bf keywords}: {\sl convection, general fluid mechanics, geophysical and geological flows}\\[5mm]

There is a flaw in the reasoning in the original paper, just before equation (2.9). Everything before that point is correct, essentially that $- \left< \boldsymbol{u} \cdot {\bf \nabla } P  \right> $ is an exact expression for dissipation in thermal convection. The conclusions obtained after that point are essentially unfounded, first that $\gamma -1 \ll 1$ is the condition of validity for the anelastic liquid approximation, second that the scaling $P' \sim K_{T0} \alpha _0 T'$ applies within the general anelastic approximation. Other conclusions are derived here, in this corrected version, on the continuity equation in the anelastic approximation.  

We first summarize the results that can be derived from the complete equations of convection. The dissipation is exactly
\begin{equation}
<\tau : \dot \epsilon>=- \left< \boldsymbol{u} \cdot {\bf \nabla } P  \right> ,
\end{equation}
 (Alboussi\`ere and Ricard 2013, equation (1.6)).
Decomposing $P = P_0 + P'$ and $\rho = \rho _0 + \rho '$, where $P_0$ and $\rho_0$ are the adiabatic, hydrostatic pressure and density, we have
\begin{eqnarray}
<\tau : \dot \epsilon> &=& - \left< \boldsymbol{u} \cdot {\bf \nabla } P_0  \right> - \left< \boldsymbol{u} \cdot {\bf \nabla } P'  \right> , \label{diss}  \\
&=& \left< \rho _0 g u_z \right>  + \left< P' {\bf \nabla} \cdot \boldsymbol{u} \right> . \label{diss2} 
\end{eqnarray}
The second line (\ref{diss2}) has been obtained using the hydrostatic equation for the reference adiabatic profile and a Gauss integration.
The dissipation appears therefore to have two contributions. The first one $\left< \rho _0 g u_z \right> $ is exactly 
$- \left< \rho ' g u_z \right> $ by global mass conservation. Because $\rho'$ and $u_z$ are generally correlated in a convective system (light material rises, dense material sinks), this term is likely positive (i.e. $<u_z>$ opposite to gravity).
The second term can be estimated using also the continuity equation  
$\partial \rho ' / \partial t + {\bf \nabla } \cdot \left[ (\rho _0 + \rho ' ) \boldsymbol{u} \right] = 0$ which implies that $\rho _ 0 {\bf \nabla } \cdot \boldsymbol{u} = - \rho ' {\bf \nabla } \cdot \boldsymbol{u} - u_z \mathrm{d} \rho _0 / \mathrm{d} z - \boldsymbol{u} \cdot {\bf \nabla} \rho ' - \partial \rho ' / \partial t$. In the limit of vanishing viscosity and thermal diffusivity, it is expected that the state variables will become close to the adiabatic hydrostatic profile, in particular $ \rho ' / \rho _0 \longrightarrow 0 $, which implies that $ \rho _ 0 {\bf \nabla } \cdot \boldsymbol{u} \approx - u_z \mathrm{d} \rho _0 / \mathrm{d} z $ at leading order. Dissipation (\ref{diss2}) can be written:  
\begin{equation}
<\tau : \dot \epsilon>  \approx - \left< \rho ' g u_z \right> - \left< P' u_z \frac{1}{\rho _0} \frac{\mathrm{d} \rho _0 }{\mathrm{d} z} \right>, \label{diss3}
\end{equation}
Then the density perturbations can be written in terms of entropy and pressure perturbations, as a linear expansion about the adiabatic profile: 
\begin{equation}
\rho' \approx - {\alpha_0\rho_0 T_0 \over c_{p0}}s'-\frac{1}{\rho _0 g} \frac{\mathrm{d} \rho _0 }{\mathrm{d} z}P', \label{rhosp}
\end{equation}
({\it cf} our equation (2.6)), leading to the more familiar
\begin{equation}
<\tau : \dot \epsilon> \approx \left<  \frac{\alpha _0 \rho _0 T_0 g}{c_{p0}} u_z s' \right> .  \label{approxdiss}
\end{equation}

If we now start from the Navier Stokes equation in the anelastic approximation
\begin{equation}
\rho _0 \frac{D \boldsymbol{u} }{D t } = - \rho _0 {\bf \nabla } \left( \frac{P'}{\rho _0 }  \right) + \frac{\alpha _0 \rho _0 T_0 g}{c_{p0}} s' \hat{\boldsymbol{e}}_z + {\bf \nabla } \cdot {\bf \tau } \label{NScorr} 
\end{equation}
 (Alboussi\`ere and Ricard 2013, equation (2.8)), we get exactly 
\begin{equation}
<\tau : \dot \epsilon> = \left<  \frac{\alpha _0 \rho _0 T_0 g}{c_{p0}} u_z s' \right> ,  \label{yyy}
\end{equation}
and using (2.6),
\begin{equation}
<\tau : \dot \epsilon> = - \left< \rho ' g u_z \right> - \left< P' u_z \frac{1}{\rho _0} \frac{\mathrm{d} \rho _0 }{\mathrm{d} z} \right>. \label{yy}
\end{equation}

The expressions of dissipation in the exact case and in the anelastic approximation, in terms of entropy perturbation, (\ref{approxdiss}) and (\ref{yyy}), or in term of density and pressure perturbations (\ref{diss3}) and (\ref{yy}) are very similar. However, and this is the weak point of our previous reasoning and the cause of this corrigendum, we used the anelastic mass conservation $\nabla \cdot \left( \rho_0 \boldsymbol{u} \right) =0$ together with the exact dissipation expression (\ref{diss2}) to conclude that  $\left< \rho _0 g u_z \right>=0$ and that dissipation in the anelastic approximation is simply $- \left< \boldsymbol{u} \cdot {\bf \nabla } P'  \right>$. In doing so, we have unduly mixed equations of two different formalisms. In the framework of anelasticity, we can only prove (\ref{yy}), and the anelastic mass conservation says nothing about $\left< \rho ' g u_z \right>$.
 Therefore, we must consider unproven that the dissipation in the anelastic
formalism can be expressed in terms of pressure variations only, i.e. we have no arguments to neglect
 the term involving $\left< \rho' u_z \right>$ in (\ref{yy})  and to retain that involving $\left< P' u_z \right>$. The rest of our paper, comparing the relative amplitudes
of terms in $<u_z s'>$, $<u_z P'>$ and $<u_z T'>$ in the anelastic approximation is most probably wrong and, in any case, not demonstrated.

 The treatment of continuity is at the origin why the expression of dissipation $ - \left< \boldsymbol{u} \cdot {\bf \nabla } P  \right> $ does not translate into $ - \left< \boldsymbol{u} \cdot {\bf \nabla } P'  \right> $ when using the anelastic approximation model. If we want that property to hold in the anelastic model, and pressure fluctuations to be a faithfull image of that in the full model, one must substitute the continuity equation ${\bf \nabla } \cdot \left[ (\rho _0 + \rho ' ) \boldsymbol{u} \right] = 0$ for the equation usually used in the anelastic model ${\bf \nabla } \cdot \left( \rho _0 \boldsymbol{u} \right) = 0 $.  
The fact that mass conservation is not well treated at first order in the perturbations from the adiabatic state can be seen from the original anelastic equations. Energy conservation (Alboussi\`ere and Ricard 2013, equation (2.4)) provides $s'$, while the momentum equation (2.8) or (\ref{NScorr}) provides $P'$ and $\boldsymbol{u}$ (with continuity (2.2)). From $s'$ and $P'$, it is possible to evaluate $\rho '$ from the linearized equation of state. Then there is no equation ensuring mass conservation, {\it i.e.} $\left< \rho ' \boldsymbol{u} \right> $ for instance is not constrained to be zero. 

In the original anelastic equations, the horizontal average of $u_z$ is zero and $\left< u_z \right> = 0$. However, the analysis in this corrigendum (in particular (\ref{diss2}) stresses the importance of the mean upward vertical velocity $\left< u_z \right> \neq 0$. Finally, the combination of that horizontally averaged vertical velocity $\overline{u_z}$ and the adiabatic hydrostatic pressure gradient produces a significant contribution to the expression of dissipation $- \left< \boldsymbol{u} \cdot {\bf \nabla } P  \right>$, of the form $\left< \overline{u_z} \rho _0 g \right>  $. That contribution is not present in the original anelastic equations and is well accounted for in a modified version of the anelastic equations when continuity is written ${\bf \nabla } \cdot \left[ (\rho _0 + \rho ' ) \boldsymbol{u} \right] = 0$.

To conclude, we have shown that the approximated continuity equation ${\bf \nabla } \cdot (\rho _0 \boldsymbol{u} ) = 0 $ is not compatible with the property that energy dissipation is equal to $ - \left< \boldsymbol{u} \cdot {\bf \nabla } P  \right> $. Hence either dissipation and/or pressure are likely to be evaluated incorrectly in the original anelastic approximation. This inconsistency is resolved when the modified continuity equation ${\bf \nabla } \cdot \left[ (\rho _0 + \rho ' ) \boldsymbol{u} \right] = 0$ is used, without affecting the anelastic nature of the approximation.

\section*{\sc References}
\noindent
{\sc 
Alboussi\`ere, T. and Ricard, Y., 2013, {\it Reflections on dissipation associated with thermal convection}, J. Fluid Mech., {\bf 725}, R1 }


\end{document}